\documentstyle[psfig]{mn}

\newif\ifAMStwofonts

\ifoldfss
  \newcommand{\rmn}[1] {{\rm #1}}
  \newcommand{\itl}[1] {{\it #1}}
  \newcommand{\bld}[1] {{\bf #1}}
  \ifCUPmtlplainloaded \else
    \NewTextAlphabet{textbfit} {cmbxti10} {}
    \NewTextAlphabet{textbfss} {cmssbx10} {}
    \NewMathAlphabet{mathbfit} {cmbxti10} {} 
    \NewMathAlphabet{mathbfss} {cmssbx10} {} 
  \fi
  \ifAMStwofonts
    \ifCUPmtlplainloaded \else
      \NewSymbolFont{upmath} {eurm10}
      \NewSymbolFont{AMSa} {msam10}
      \NewMathSymbol{\upi}     {0}{upmath}{19}
      \NewMathSymbol{\umu}     {0}{upmath}{16}
      \NewMathSymbol{\upartial}{0}{upmath}{40}
      \NewMathSymbol{\leqslant}{3}{AMSa}{36}
      \NewMathSymbol{\geqslant}{3}{AMSa}{3E}
      \let\oldle=\le     \let\oldleq=\leq
      \let\oldge=\ge     \let\oldgeq=\geq
      \let\leq=\leqslant \let\le=\leqslant
      \let\geq=\geqslant \let\ge=\geqslant
    \fi
  \fi
\fi 

\ifnfssone
  \newmathalphabet{\mathit}
  \addtoversion{normal}{\mathit}{cmr}{m}{it}
  \addtoversion{bold}{\mathit}{cmr}{bx}{it}
  \newcommand{\rmn}[1] {\mathrm{#1}}
  \newcommand{\itl}[1] {\mathit{#1}}
  \newcommand{\bld}[1] {\mathbf{#1}}
  \def\textbfit{\protect\txtbfit}
  \def\textbfss{\protect\txtbfss}
  \long\def\txtbfit#1{{\fontfamily{cmr}\fontseries{bx}\fontshape{it}%
    \selectfont #1}}
  \long\def\txtbfss#1{{\fontfamily{cmss}\fontseries{bx}\fontshape{n}%
    \selectfont #1}}
  \newmathalphabet{\mathbfit} 
  \addtoversion{normal}{\mathbfit}{cmr}{bx}{it}
  \addtoversion{bold}{\mathbfit}{cmr}{bx}{it}
  \newmathalphabet{\mathbfss} 
  \addtoversion{normal}{\mathbfss}{cmss}{bx}{n}
  \addtoversion{bold}{\mathbfss}{cmss}{bx}{n}
  \ifAMStwofonts
    \ifCUPmtlplainloaded \else
      \UseAMStwoboldmath
      \makeatletter
      \new@mathgroup\upmath@group
      \define@mathgroup\mv@normal\upmath@group{eur}{m}{n}
      \define@mathgroup\mv@bold\upmath@group{eur}{b}{n}
      \edef\UPM{\hexnumber\upmath@group}
      \new@mathgroup\amsa@group
      \define@mathgroup\mv@normal\amsa@group{msa}{m}{n}
      \define@mathgroup\mv@bold\amsa@group{msa}{m}{n}
      \edef\AMSa{\hexnumber\amsa@group}
      \makeatother
      \mathchardef\upi="0\UPM19
      \mathchardef\umu="0\UPM16
      \mathchardef\upartial="0\UPM40
      \mathchardef\leqslant="3\AMSa36
      \mathchardef\geqslant="3\AMSa3E
      \let\oldle=\le     \let\oldleq=\leq
      \let\oldge=\ge     \let\oldgeq=\geq
      \let\leq=\leqslant \let\le=\leqslant
      \let\geq=\geqslant \let\ge=\geqslant
    \fi
  \fi
\fi 

\ifnfsstwo
  \newcommand{\rmn}[1] {\mathrm{#1}}
  \newcommand{\itl}[1] {\mathit{#1}}
  \newcommand{\bld}[1] {\mathbf{#1}}
  \def\textbfit{\protect\txtbfit}
  \def\textbfss{\protect\txtbfss}
  \long\def\txtbfit#1{{\fontfamily{cmr}\fontseries{bx}\fontshape{it}%
    \selectfont #1}}
  \long\def\txtbfss#1{{\fontfamily{cmss}\fontseries{bx}\fontshape{n}%
    \selectfont #1}}
  \DeclareMathAlphabet{\mathbfit}{OT1}{cmr}{bx}{it}
  \SetMathAlphabet\mathbfit{bold}{OT1}{cmr}{bx}{it}
  \DeclareMathAlphabet{\mathbfss}{OT1}{cmss}{bx}{n}
  \SetMathAlphabet\mathbfss{bold}{OT1}{cmss}{bx}{n}
  \ifAMStwofonts
    \ifCUPmtlplainloaded \else
      \DeclareSymbolFont{UPM}{U}{eur}{m}{n}
      \SetSymbolFont{UPM}{bold}{U}{eur}{b}{n}
      \DeclareSymbolFont{AMSa}{U}{msa}{m}{n}
      \DeclareMathSymbol{\upi}{0}{UPM}{"19}
      \DeclareMathSymbol{\umu}{0}{UPM}{"16}
      \DeclareMathSymbol{\upartial}{0}{UPM}{"40}
      \DeclareMathSymbol{\leqslant}{3}{AMSa}{"36}
      \DeclareMathSymbol{\geqslant}{3}{AMSa}{"3E}
      \let\oldle=\le     \let\oldleq=\leq
      \let\oldge=\ge     \let\oldgeq=\geq
      \let\leq=\leqslant \let\le=\leqslant
      \let\geq=\geqslant \let\ge=\geqslant
    \fi
  \fi
\fi 

\ifCUPmtlplainloaded \else
  \ifAMStwofonts \else 
    \def\upi{\pi}
    \def\umu{\mu}
    \def\upartial{\partial}
  \fi
\fi

\title{Pulsational $M_V$ versus $[Fe/H]$ relation(s) for globular cluster 
RR Lyrae variables }
\author[F. Caputo et al.]
       {F. Caputo$^1$, V. Castellani$^2$, M. Marconi$^3$ and V. Ripepi$^3$
\\
        $^1$ Osservatorio Astronomico di Roma, via di Frascati 33, 
00040 Monte Porzio Catone, Italy\\
	$^2$ Dipartimento di Fisica, Piaza Torricelli 2, 56100 Pisa, 
Italy\\            
        $^3$ Osservatorio Astronomico di Capodimonte, 
Via Moiariello 16, 80131 Napoli, Italy}
\date{}

\pagerange{\pageref{firstpage}--\pageref{lastpage}}
\pubyear{1994}

\begin{document}

\maketitle

\label{firstpage}

\begin{abstract}

We  use the results from recent computations of 
updated non-linear convective pulsating models to constrain
the distance modulus of Galactic globular clusters through
the observed periods of first overtone ($RR_c$) pulsators.
The resulting  relation between the mean absolute magnitude of 
RR Lyrae stars $<M_V(RR)>$ and the heavy element content $[Fe/H]$ 
appears well in the
range of several previous empirical calibrations, but with
a non linear dependence on $[Fe/H]$ so that the slope of 
the relation increases when moving towards   
larger metallicities. 
On this ground, our results suggest that metal-poor ($[Fe/H]<$-1.5) and 
metal-rich ($[Fe/H]>$-1.5) variables follow two 
different linear $<M_V(RR)>-[Fe/H]$ relations. 
Application to RR Lyrae stars in the metal-poor globular clusters
of the Large Magellanic Cloud 
provides a LMC distance modulus of the
order of 18.6 mag, thus supporting  the ``long" distance scale.  
The comparison with recent predictions based on updated stellar evolution 
theory is shortly
presented and discussed.

\end{abstract}


\section{Introduction}

The intrinsic luminosity of RR Lyrae variables has been for
a long time a very popular way  to give reasonable estimates
of the distance to globular clusters (GCs) both in the Milky Way and in 
Local Group galaxies (Magellanic Clouds, M31) and, in turn, to constrain 
the age of these very old stellar systems. However, 
notwithstanding the large body of work, a general consensus on
a precise evaluation of such a luminosity has been not yet achieved.
One may notice that a firm knowledge of RR Lyrae luminosities
would be of paramount relevance, since it  would provide an
independent test of the Cepheid distance scale as well as a 
reliable calibration of several secondary distance indicators 
(as, e.g., the GC luminosity function or the  Red Giant Tip) 
for external galaxies, thus providing important clues on 
the value of the Hubble constant $H_0$. On these grounds, RR Lyrae variables 
could represent relevant milestones on the path to set both a lower 
and an upper limit to the age of the Universe, playing a fundamental
role in several astrophysical problems ranging 
from stellar evolution to cosmological models.
 
From the observational side, studies dealing with the absolute 
magnitude $M_V(RR)$ of RR Lyraes and with the dependence of these 
magnitudes on the heavy 
element content $[Fe/H]$ has yielded to the well known debate  
between the so-named "short" and "long" distance scales. As recently 
reviewed by Cacciari (1999), empirical  estimates of $M_V(RR)$ 
for RR Lyrae stars at $[Fe/H]$=-1.6  actually range from about 0.4 mag  
to 0.7 mag, thus leaving an uncertainty of $\sim\pm$ 0.2  mag on the 
derived distance moduli (see also Popowski \& Gould 1999). 

Different estimates have been also given
for the dependence of these magnitudes on the star metallicity.
As a matter of the fact, for the often assumed 
linear relation 

$$<M_V(RR)>=a+b[Fe/H]$$ 

\noindent
one finds in the literature evaluations of the coefficient $"b"$ 
{\bf mainly in the range} $b\sim 0.18\pm 0.03$ to $\sim$0.30, where the 
former value is based on the 
Baade-Wesselink method (see, e.g., Fernley et al. 1998b [Fn98b]) 
and the latter
value has been early suggested by Sandage (1993 [Sa93]) when discussing the 
period-metallicity  relation for field and GC RR Lyrae pulsators. 
{\bf However, an even milder slope has been suggested by Fusi Pecci
et al. (1996), who investigated eight globular clusters in M31 to
derive, over the range -1.8$<{[Fe/H]}<$-0.4, $$<M_V(HB)>=(0.13\pm0.07)*[Fe/H]+(0.95\pm0.09)$$}. 

The recent release of HIPPARCOS statistical and trigonometric parallaxes 
for halo RR Lyraes ($[Fe/H]\le$-1.30) has not clarified the issue: 
one may indeed recall that Fernley et al. 1998a [Fn98a] and Groenewegen \& 
Salaris (1999 [GS99]), both assuming the same slope $b$=0.18, 
give a zero-point 
of $1.05\pm0.15$ mag and 0.77$\pm0.26$ mag, respectively, as derived from an
identical sample of variables but using different approaches (statistical 
parallaxes or reduced parallaxes, respectively). In the meantime, 
McNamara 1999 [MN99] claims that Baade-Wesselink results for 
variables with $[Fe/H]>-1.5$ yield a quite different relation  
as given by 

$$<M_V(RR)>=1.06+0.32[Fe/H] \label{[MN99]}$$ 

On the theoretical side, the literature already contains several  
sets of horizontal branch (HB) evolutionary models computed for 
wide ranges of the overall metallicity ($Z$ in the range 
of 0.0001 to 0.02) which 
provide the "theoretical route" to the calibration of the $M_V(RR)$ 
versus $[Fe/H]$ relation. One finds that almost all the recent 
theoretical predictions concerning the absolute magnitude 
$M_V(ZAHB)$ of the 
zero age horizontal branch (ZAHB) sequence at the RR Lyrae instability 
strip confirm the non-linear dependence of $M_V(ZAHB)$ on $\log Z$ formerly
suggested by Castellani, Chieffi \& Pulone (1991 [CCP]). 
However, the scaling of the overall metallicity  $Z$ to the 
measured  $[Fe/H]$ values 
could be a tricky matter since the classical assumption of solar-scaled 
chemical mixtures is likely inappropriate to GC stars.  
There is indeed a growing 
observational evidence for a significant enhancement 
of $\alpha$-elements with respect 
to iron ($[\alpha/Fe] \sim 0.3$) in GC and field metal-poor stars 
(see Carney et al. 1997, Gratton et al. 1997). Moreover, one 
has to bear in mind that observed RR Lyrae samples do contain stars 
evolved off their original ZAHB position. Thus, realistic predictions on 
the average magnitude 
$<M_V(RR)>$ require the evaluation of the evolutionary effects, possibly 
through synthetic HB  simulations (SHB).

The wide grids of SHBs so far published (e.g., Lee, Demarque \& Zinn 1990, 
Lee 1991, Bencivenni et al. 1991, Caputo et al. 1993) have already shown  
that the predicted mean magnitude of RR Lyrae stars $<M_V(RR)>$ significantly 
depends, with everything else
being constant, on the  HB morphology. Simulations based on slightly modified
CCP  models yielded Caputo (1997) to  
suggest  

$$<M_V(RR)>=1.19+0.19 \log Z$$
\noindent
for RR Lyrae-rich metal-poor GCs 
with $\log Z\le -3.0$, whereas for larger metallicities the theory 
gives 

$$<M_V(RR)>=1.57+0.32 \log Z$$

\noindent
Similar results have been more recently found by Demarque et al. 
(1999), who definitively reject the existence of a unique linear 
relation covering the metallicities spanned 
by GCs, 
confirming that the slope of the predicted 
$M_V(RR)-\log Z$ relation depends on the 
metallicity range and that the HB morphology of each cluster must be 
taken in the due account when using RR Lyrae stars as distance indicators.

On these grounds, one is tempted to conclude that theoretical 
and observational investigations do show a sort of consistency: the former 
give warnings against a "universal" linear $M_V(RR)-[Fe/H]$ 
relation, the latter fail to reach an agreement on both its slope 
and zero-point! 

Within such a confusing scenario, one has to mark the
seminal attempts made by Sandage (Sa93 and references therein) 
to use RR Lyrae periods to constrain the luminosity of these stars. 
This appears 
a quite relevant approach, since periods are firm and safe observational 
parameters, independent of distance and reddening. To discuss Sandage's
philosophy, one has to recall that since the  pioneering work by Christy (1966) 
and Stellingwerf (1975, 1984) pulsating models have 
suggested the existence within the 
instability strip of a region where both  fundamental ($RR_{ab}$) and first 
overtone ($RR_c$) modes are stable (see Bono \& Stellingwerf 1994, Bono et al. 
1997a, Bono et al. 1997c). The boundaries of this "either-or" 
region, namely the  
fundamental blue edge (FBE) at the higher temperature side  and the 
first overtone red edge (FORE) at the lower temperature side, encompass
for each given luminosity  the range of temperatures (or colours) 
where the mode-shift (i.e. the 
transition from $RR_{ab}$ to $RR_c$) may occur.  
Assuming that for both Oosterhoff type I 
(OoI) and Oosterhoff type II (OoII) globular clusters 
this transition occurs 
at the blue edge for fundamental pulsation and using periods and $B-V$ colours 
of the shortest period $RR_{ab}$  in clusters and in the field, 
Sa93 derives the star
luminosity from the well established period-mass-luminosity-temperature
relation. In this way he obtains the relation

$$M_V(RR)=0.94+0.30[Fe/H] \label{[Sa93]}$$

\noindent
which accounts for the Oosterhoff dichotomy in Galactic
globular clusters as mainly due to a luminosity effect. 
However, a re-analysis by 
Fernley (1993 [Fn93]), using $V-K$ colours and a limited sample 
of clusters with low and well-known reddening, yields

$$M_V(RR)=0.84+0.19[Fe/H] \label{[Fn93]}$$

More recently the assumption of a unique $RR_{ab}/RR_c$ 
transition line has been 
questioned by Bono, Caputo \& Marconi (1995), who concluded that the Oosterhoff 
dichotomy is largely the result of different transition lines 
in OoI (near FBE) and OoII (near FORE) clusters, as early suggested by 
van Albada and Baker (1973).
However, the pulsation theory predicts the limits of the whole instability 
strip, 
as given by the first overtone blue edge (FOBE) and the 
fundamental red edge (FRE), 
without any ambiguity about the actual pulsation mode. 
On this basis, Caputo (1997)
used a preliminary set of pulsating models to show  
that theoretical predictions 
on the pulsator distribution in the 
period-absolute magnitude $M_V-\log P$ plane 
can constrain the distance to RR Lyrae-rich globular clusters. 

In recent times, the RR Lyrae pulsating models have been updated and 
extended to wide ranges of mass and chemical composition, 
shedding light on the dependence of the instability strip on the 
metal content. In this paper we will take advantage of these improvements
to reconsider the Caputo (1997) analysis. The updated $M_V-\log P$ 
relations at FOBE and FRE 
are discussed in the following Sect. 2, while  Sect. 3 presents the comparison 
with observation and the derived 
"pulsational" distance moduli for a selected
sample of well-studied GCs. The resulting dependence of 
our $<M_V(RR)>$ values on the cluster  
metallicity is discussed in Sec. 4 
in comparison with both empirical relations and 
recent theoretical HB models. 
Some concluding remarks will close the paper. 

\section{The predicted $M_V-\log P$ diagram}

\begin{figure*}
\vspace{4cm}
\caption{ Globular cluster RR Lyraes in the $M_V-\log P$ plane. Each group 
of figures represents a ``class'' of metallicity. 
In each panel filled and open 
circles are $RR_c$ and $RR_{ab}$ pulsators, respectively, while the solid 
lines are the theoretical boundaries of the instability strip 
(see text for details).  
The labelled apparent distance moduli are obtained by constraining 
the observed $RR_c$ distribution to match the predicted blue limit of the 
pulsation region, under the assumption of solar-scaled chemical composition. \label{fig1}}
\end{figure*}

The non-linear convective hydrodinamical code used for pulsating 
models has been already presented in a series of papers 
(Bono \& Stellingwerf 1994, Bono, Caputo \& Marconi 1995, Bono et al. 1997c) 
and it will be not further discussed. With respect to previous 
computations, the new models differ in the adopted opacity tables, using
the most updated compilations by 
Iglesias \& Rogers (1996) and extending in such a way
the preliminary results presented
in Caputo et al. (1999) for  $Z$=0.001 and $M=0.65 M_{\odot}$.

Table 1 presents temperatures, absolute magnitudes and periods of stars
at FOBE or FRE, for selected choices on $Z$ and suitable 
values of the star mass ($M$), luminosity ($L$) and helium abundance ($Y$).
Absolute magnitudes are derived using the bolometric 
corrections provided by Castelli, Gratton \& Kurucz (1997a,b). 
The adopted $Y$-values reasonably account for the extra helium brought to the 
stellar surface by the first dredge-up as well as for a galactic 
enrichment as given by $\Delta Y / \Delta Z \sim 2.5$.
As discussed in Caputo, Marconi \& Santolamazza (1998), mild variations 
of $Y$, as due also to uncertainties in the efficiency of
element sedimentation (see Cassisi et al. 1998, 1999 [Cs99]), have negligible
effects on the temperature of the instability edges and, in turn, on the 
related pulsational periods. Stellar masses and luminosities have been 
chosen in such a way to reasonably encompass available expectations
about these evolutionary parameters for GC RR Lyrae pulsators.

\begin{table*}
 \centering
 \caption[]{Effective temperature, period and absolute magnitude 
 at FOBE and FRE for the various chemical 
compositions, masses and luminosities. \label{tab1}}
 \begin{tabular}{cccccccccc}
 $Z$  &   $Y$  &   $M/M_{\odot}$ &  $\log L/ \log L_{\odot}$ & $T_e(FOBE)$ & $T_e(FRE)$ 
   & $M_V(FOBE)$& $\log P(FOBE)$ & $M_V(FRE)$ & $\log P(FRE)$ \\
 0.0001   & 0.24    & 0.80   &      1.81   &    7150    &    5950 &     0.231   &     -0.4606   &        0.350    &   -0.0691   \\
 0.0001   & 0.24    & 0.80   &      1.72   &    7250    &    6050 &     0.456   &     -0.5535   &       0.565    &    -0.1695   \\
 0.0001   & 0.24    & 0.75   &      1.81   &    7150    &    5850 &     0.229   &     -0.4439   &       0.361    &   -0.0250    \\
 0.0001   & 0.24    & 0.75   &      1.72   &    7250    &    5950 &     0.454   &     -0.5367   &       0.577    &    -0.1258   \\
 0.0004   & 0.24    & 0.70   &      1.81   &    7050    &    5800 &     0.221   &     -0.4026   &        0.350    &    0.0123    \\
 0.0004   & 0.24    & 0.70   &      1.72   &    7150    &    5900 &     0.446   &     -0.4957   &       0.565    &   -0.0887     \\
 0.0004   & 0.24    & 0.70   &      1.61   &    7350    &    6000 &     0.718   &     -0.6245   &       0.831    &    -0.2061   \\
 0.0004   & 0.24    & 0.65   &      1.81   &    7050    &    5800 &     0.219   &     -0.3834   &       0.349    &    0.0342    \\
 0.0004   & 0.24    & 0.65   &      1.72   &    7150    &    5900 &     0.444   &     -0.4765   &       0.563    &   -0.0668     \\
 0.0004   & 0.24    & 0.65   &      1.61   &    7350    &    6000 &     0.716   &     -0.6053   &        0.83    &    -0.1842   \\
 0.001    & 0.24    & 0.65   &      1.72   &    7150    &    5850 &     0.432   &     -0.4745   &       0.553    &   -0.0514    \\
 0.001    & 0.24    & 0.65   &      1.61   &    7250    &    5950 &     0.709   &     -0.5835   &       0.819    &    -0.1691   \\
 0.001    & 0.24    & 0.60   &      1.72   &    7150    &    5850 &      0.430   &     -0.4537   &       0.552    &   -0.0278    \\
 0.001    & 0.24    & 0.60   &      1.61   &    7250    &    5950 &     0.707   &     -0.5627   &       0.819    &    -0.1455   \\
 0.002    & 0.24    & 0.62   &      1.65   &    7250    &    5900 &     0.592   &     -0.5374   &       0.705    &    -0.1069   \\
 0.002    & 0.24    & 0.62   &      1.55   &    7350    &    6000 &     0.844   &     -0.6381   &       0.946    &    -0.2159   \\
 0.002    & 0.24    & 0.57   &      1.65   &    7200    &    5850 &     0.593   &     -0.5056   &       0.711    &   -0.0695  \\
 0.002    & 0.24    & 0.57   &      1.55   &    7300    &    5950 &     0.844   &     -0.6064   &       0.952    &    -0.1787   \\
 0.006    & 0.255   & 0.58   &      1.55   &    7250    &    5850 &     0.823   &     -0.5986   &       0.935    &    -0.1554   \\
 0.01     & 0.255   & 0.58   &      1.65   &    7050    &    5650 &     0.565   &     -0.4762   &       0.704    &    -0.0132   \\ 
 0.01     & 0.255   & 0.58   &      1.57   &    7150    &    5750 &     0.764   &     -0.5613   &       0.888    &    -0.1065     \\
 0.01     & 0.255   & 0.58   &      1.51   &    7150    &    5850 &     0.917   &     -0.6098   &       1.025    &    -0.1825    \\
 0.02     & 0.28    & 0.53   &      1.81   &    6750    &    5350 &     0.149   &     -0.2591   &       0.353    &     0.2369   \\
 0.02     & 0.28    & 0.53   &      1.61   &    7050    &    5850 &     0.641   &     -0.4837   &       0.785    &   -0.0121    \\
 0.02     & 0.28    & 0.53   &      1.51   &    7150    &    5650 &     0.891   &     -0.5849   &       1.002    &    -0.1477   \\
 0.02     & 0.28    & 0.53   &      1.41   &    7350    &    5950 &     1.139   &     -0.7056   &       1.241    &    -0.2569   \\
 0.02     & 0.28    & 0.53   &      1.21   &    7550    &    6150 &     1.645   &     -0.9061   &       1.719    &    -0.4741  
 \end{tabular}
 \end{table*}

It should be noted that, for each given set of entry parameters, 
the  computations 
have been performed by steps of 100 K and that we adopt as limits of the 
instability region the average effective temperature between the last 
pulsating model and the first non pulsating one. It follows that the 
intrinsic uncertainty of the FOBE and FRE temperatures in Table 1
is $\pm$ 50 K, which in terms of period means 
$\delta \log P\sim \pm$0.01 (see Caputo et al. 1998). From 
the data in Table 1 one derives analytical expressions  
connecting the absolute V-magnitude at the instability edges with the 
pulsator 
period, mass and metallicity, as given by

\begin{eqnarray} 
M_V(FOBE)= & -0.685-2.255 \log P(FOBE)+ \nonumber \\ 
& -1.259 \log M/M_{\odot}+0.058 \log Z \label{eq1}  
\end{eqnarray}
\begin{eqnarray} 
M_V(FRE)= & +0.552-2.018 \log P(FRE)+ \nonumber \\ 
& -1.348\log M/M_{\odot}+0.108 \log Z, \label{eq2}    
\end{eqnarray}

\noindent
with a rms scatter $\sigma_V$=0.027 mag. 

However, a further source of uncertainty  on the predicted pulsation edges    
is given by the efficiency of convection in the external layers. The lack of a 
rigorous treatment of superadiabatic convection 
is indeed a well known  fault in the whole stellar evolution theory 
and almost all the evolutionary sequences are calculated within 
the so-called "mixing-length" scenario which involves an adjustable
parameter $l/H_p$, the ratio of the mixing-length to the pressure scale height. 
Our pulsating models are consistent with this scenario and, in 
order to close the system of convective and dynamic equations, adopt 
$l/H_p=1.5$, in reasonable agreement with 
the values generally used for evolutionary 
computations. Since the effect of convection is to quench 
pulsation, variations of  $l/H_p$ lead to variations in the effective 
temperature of the boundaries for instability, with the amount 
of this effect decreasing from the red to the blue edge.
 
To have light on such an uncertainty, we performed 
suitable numerical experiments, 
finding that  decreasing $l/H_p$ down to 1.1 (i.e. decreasing the 
efficiency of convection and thus increasing the local temperature gradients)
the FOBE periods decrease by $\delta \log P \sim 0.029$, while with $l/H_p$=2.0
these periods increase by $\delta \log P\sim 0.017$. On this ground 
one can estimate that mixing-length values in the range $l/H_p=1.3-1.8$, 
as widely adopted in the relevant literature, 
yield an additional uncertainty on FOBE periods  
by $\delta \log P\sim\pm 0.01$. Since at the red side of the instability 
strip  the mixing-length affects much more significantly the predicted periods,  
in the following we will rely on 
theoretical predictions
concerning FOBE only, temptatively putting the red edge of 
the pulsation region at 
$\Delta \log P=0.45$ with respect to FOBE. 

\begin{table}
\centering
\caption[]{Masses of $ZAHB$ models with $\log T_e=3.85$ and ages in the 
range of 8 to 18 Gyrs (from Bono et al. 1997b).  
\label{tab2}}
\begin{tabular}{cc}
$Z$  &      $M/M_{\odot}$ \\
0.0001 &  0.796    \\
0.0003 &  0.711    \\
0.0006 &  0.671    \\
0.001  &  0.648    \\
0.003  &  0.608    \\
0.006  &  0.585    \\
0.01   &  0.575    \\
0.02   &  0.545    \\
\end{tabular}
\end{table}

As a relevant point, when constraining the luminosities
of ZAHB pulsators in the range  covered by current evolutionary predictions 
for various metallicities and for ages in the range from  8 to 18 Gyr,
one finds that  data in Table 1 predict FOBE effective
temperatures very close to $\log T_e=3.85$, without significant
variation with the metal content. This constant value is due to the
balancing effects of metallicity on the FOBE temperature 
(which decreases if increasing only the luminosity) and 
ZAHB luminosity: when decreasing $Z$ only, 
the FOBE would become hotter, but the contemporary increase of   
the ZAHB luminosity eventually leaves  unchanged the effective temperature.
More in general, we can use such an evidence to safely take from evolutionary 
theories the predicted masses
for HB pulsators at the blue side of the instability strip. 
Luckily enough, at variance with luminosities, the evolutionary masses 
have passed substantially unchanged the many
improvements affecting HB models in the last years, 
thus representing a rather firm and trustworthy prediction. 
{\bf As a relevant point, one finds that such evolutionary prediction
appears in close agreement with independent mass estimates from the
period ratios of double-mode RR Lyrae pulsators (see Cox 1991,  Bono
et al. 1996).}\par

Thus, by inserting  into eq. (1) the predicted mass of the ZAHB model 
at $\log T_e=3.85$  
as presented in Table 2 (from Bono et al. 1997b), one finally gets 
the period-luminosity-metallicity relation for {\it evolutionary} 
FOBE pulsators, as given by
\begin{eqnarray} 
M_V(FOBE)= & -0.178-2.255 \log P(FOBE)+ \nonumber \\ 
 & +0.151 \log Z,   \label{eq3}
\end{eqnarray}
\noindent
with a total intrinsic dispersion 
(including the above uncertainty of $\pm$50 K, the  
mixing-length effects and mass 
variations by 5\% the values in Table 2) of $\sigma_V=0.065$ mag.

Here one should note that the mass of HB models does depend on the
abundance of $\alpha$-elements. However,  one can benefit by 
the {\it principle of correspondence} 
for which the evolutionary behaviour of HB stars (and the predicted
pulsator masses) depends 
on the overall metallicity $Z$, independently of the internal 
ratio between $\alpha$- and heavy elements (see, e.g., Bencivenni et al. 
1991, Salaris, Chieffi \& Straniero 1993).   
In other words, the comparison between the predicted  
period-luminosity-metallicity relation given in eq. (3)  
and the GC RR Lyrae distribution in the observed $V-\log P$ plane 
only requires  that the scaling between the measured $[Fe/H]$ and the 
overall metallicity $Z$ is properly evaluated as 
\begin{equation}
\log Z=[Fe/H]-1.61+\log (0.638f+0.362),   \label{eq4}
\end{equation}
where $f$ is the $\alpha$-enhancement factor with respect 
to iron (Salaris et al. 1993). 

\begin{table*}
\centering
\caption[]{Observed parameters for the selected globular clusters. \label{tab3}}
\begin{tabular}{lccccc}
  Name	&	$[Fe/H]$ &  $HB$   & $<V_{RR}>$  & $\sigma_{<V_{RR}>}$ & Ref. (RR Lyrae) \\
NGC1851	        & -1.05  &  -0.36  &    16.03  & 0.09  & Walker (1998)	\\
NGC4147		& -1.54	 &  +0.55  &    16.98  & 0.08  & Newburn (1957)  \\
IC4499		& -1.26  &  +0.11  &    17.64  & 0.08  & Walker \& Nemec (1996) \\
NGC4590 M68	& -1.99	 &  +0.17  &    15.64  & 0.04  & Walker (1994)    \\
NGC4833		& -1.58  &  +0.93  &    15.33  & 0.15  & Demers \& Wehlau (1977) \\
NGC5053		& -1.98  &  +0.52  &    16.64  & 0.11  & Nemec et al. (1995)   \\
NGC5272 M3	& -1.34  &  +0.08  &    15.63  & 0.10  & Carretta et al. (1998) \\
NGC5466		& -2.14  &  +0.58  &    16.46  & 0.07  & Corwin, Carney \& Nifong (1999) \\
NGC5904 M5	& -1.11  &  +0.31  &    15.06  & 0.08  & Caputo et al. (1999) \\
NGC6171 M107	& -0.91  &  -0.73  &    15.60  & 0.10  & Dickens (1970) \\
NGC6333 M9	& -1.56  &  +0.87  &    16.26  & 0.07  & Clement \& Shelton (1999) \\
NGC6362		& -0.96	 &  -0.58  &    15.25  & 0.07  & Walker (1999) \\
NGC6809 M55	& -1.61  &   0.87  &    14.37  & 0.06  & Olech et al. (1999) \\
NGC7006		& -1.35  &  -0.28  &    18.79  & 0.14  & Wehlau, Slawson \& Nemec (1999) \\
NGC7078 M15	& -2.12  &  +0.67  &    15.83  & 0.06  & Bingham et al. (1984) \\
NGC7089 M2	& -1.31  &  +0.96  &    15.97  & 0.09  & Lee \& Carney (1999)
\end{tabular}
\end{table*}

\section{The observed $V-\log P$ diagram}

Since our analysis will be  focused on the predicted FOBE, 
we selected well-studied 
clusters with statistically significant numbers of $RR_c$ stars.
The sample of the used GCs is presented in Table 3, 
which gives for each cluster 
the adopted iron content $[Fe/H]$ (from Carretta \& Gratton (1997) or from
Zinn \& West (1984) and Rutledge, Hesser \& Stetson 
(1997) values transformed into the Carretta \& Gratton 
metallicity scale), HB type (Harris 1996) and mean  
visual magnitude $<V_{RR}>$ of RR Lyrae stars. 
 
For each assumption about the
globular cluster distance modulus one 
obtains the 
distribution of the cluster RR Lyraes in the $M_V-\log P$ plane. 
We derive 
a ``pulsational" evaluation of the cluster 
distance modulus 
by constraining the observed $RR_c$ distribution to match 
the predicted blue limit of the pulsation region in order to have 
no variables in the hot stable region. 
Figure 1 shows the result of such a procedure, 
by assuming for the cluster sample 
a solar-scaled chemical composition, i.e. $f$=1 in eq. (4). 
As already stated, our analysis is focused on 
$RR_c$ stars and the right edge of the instability 
strip has been simply placed at 
$\Delta \log P=0.45$ with respect to the left edge. However, it 
seems worthy of notice the fair agreement found also between the 
predicted FRE and the $RR_{ab}$ distribution.
 
The derived GC apparent distance moduli $DM_V$ are summarised in Table 4, 
together with 
the resulting mean absolute magnitude 
$<M_V(RR)>$ of RR Lyrae stars. 
The total errors on $<M_V(RR)>$ listed in Table 4 account 
for the observed dispersion 
$\sigma_{<V(RR>}$ (see Table 3) and   
the predicted total uncertainty ($\pm$0.07 mag) of the FOBE 
period-luminosity-metallicity relation. The last column gives the 
weight $W$ to our $<M_V(RR)>$ estimates, as simply derived from the 
number of $RR_c$ stars matching the predicted FOBE.  

\begin{table}
\centering
\caption[]{Distance moduli ($\pm 0.07$ mag) and RR Lyrae 
mean absolute magnitudes for 
solar-scaled chemical compositions. The asterisks mark 
OoII globular clusters. \label{tab4}}
\begin{tabular}{lccccc}
Name              &$[Fe/H]$  &  $HB$   & $DM_V$  &   $M_V(RR)$     & $W$ \\
NGC1851	          &   -1.05  &  -0.36  & 15.40   &   0.63$\pm$0.11 & 3   \\
NGC4147           &   -1.54  &   0.55  & 16.42   &   0.56$\pm$0.10 & 2   \\ 
IC4499            &   -1.26  &   0.11  & 17.02   &   0.62$\pm$0.10 & 2   \\
NGC4590   M68     &   -1.99  &   0.17  & 15.20   &   0.44$\pm$0.08 & 1*   \\
NGC4833           &   -1.58  &   0.93  & 14.95   &   0.38$\pm$0.16 & 1*   \\
NGC5053           &   -1.98  &   0.52  & 16.31   &   0.33$\pm$0.13 & 1*   \\
NGC5272   M3      &   -1.34  &   0.08  & 15.00   &   0.63$\pm$0.12 & 2   \\
NGC5466           &   -2.14  &   0.58  & 16.07   &   0.39$\pm$0.10 & 2*   \\
NGC5904   M5      &   -1.11  &   0.31  & 14.37   &   0.69$\pm$0.10 & 3   \\
NGC6171   M107    &   -0.91  &  -0.73  & 14.81   &   0.79$\pm$0.12 & 1   \\
NGC6333   M9      &   -1.56  &   0.87  & 15.80   &   0.46$\pm$0.10 & 3   \\
NGC6362           &   -0.96  &  -0.58  & 14.54   &   0.71$\pm$0.10 & 4   \\
NGC6809   M55     &   -1.61  &   0.87  & 13.95   &   0.42$\pm$0.09 & 3*   \\
NGC7006           &   -1.35  &  -0.28  & 18.19   &   0.60$\pm$0.16 & 2   \\
NGC7078   M15     &   -2.12  &   0.67  & 15.48   &   0.35$\pm$0.09 & 4*   \\
NGC7089   M2      &   -1.31  &   0.96  & 15.45   &   0.52$\pm$0.11 & 4*
\end{tabular}
\end{table}

Note that the results in Table 2 refer to solar-scaled chemical compositions. 
If the chemical mixtures are  $\alpha$-enhanced, then for each cluster
the nominal metallicity Z increases [see eq. (4)] and the derived distance modulus
decreases [see eq. (3)]. As a matter of example, with $f$=3  
all the distance moduli in 
Table 4 have to be decreased by 0.05 mag, with a consequent increase 
of $<M_V(RR)>$.

\section{The $<M_V(RR)>$ versus metallicity relation(s)}

The final correlation between our $<M_V(RR)>$ and the cluster metallicity  
$[Fe/H]$ is presented 
in Fig. 2 for the two cases $f$=1 and $f$=3, assuming for  
each $[Fe/H]$ an error of $\pm$ 0.15 dex. The same figure
shows the already quoted observational calibrations based on 
RR Lyrae periods (Sa93, Fn93), HIPPARCOS data for field RR Lyraes (Fn98a) 
and the Baade-Wesselink method (Fn98b).  
We add the result by Carretta et al. (1999 [Cr99]), as based on 
HIPPARCOS parallaxes for field subdwarfs  and 
Main-Sequence fitting procedure, while the GS99 relation, 
which is only 0.03 mag fainter than 
Cr99, is not presented for the sake of clearness. 

Inspection of Fig. 2 reveals that there is a general
agreement between pulsational and other empirical calibrations, 
except the Fn98a relation which 
definitively suggests too faint magnitudes. 
If $f$=1, then also the Fn98b relation appears fainter with 
respect to our results. 
However, one derives that 
none of the empirical linear calibrations is able to fully 
match our pulsational results over the whole 
range of metal content. The lack of a full agreement  is largely  due
to the fact that our data foresee a non linear relation between
$<M_V(RR)>$ and metallicity, in agreement with stellar evolution 
theoretical predictions. 
We show in Table 5 that a bare linear best fit to the data in Table 4, 
starting from the four metal-poorest clusters with 
$[Fe/H]\sim$-2.0 (NGC 4590, NGC 5053, NGC 5466, NGC 7078) 
and regularly increasing the metallicity range, yields a 
$M_V(RR)-[Fe/H]$ relation which becomes steeper and 
steeper when moving towards metal-rich clusters, suggesting 
a change in slope at $[Fe/H]\sim$-1.5. It seems worthy of 
notice that this result agrees 
with the recent analysis of RR Lyrae variables in the 
field (MN99) and the globular cluster  
$\omega$ Centauri 
(Rey et al. 2000).

\begin{table}
\centering
\caption[]{The slope of the $M_V(RR)-[Fe/H]$ relation as a function of the 
metallicity range. 
\label{tab5}}
\begin{tabular}{cc}
$[Fe/H]$-range  &   $b$ \\
-2.00/-1.60 &	  0.05$\pm$0.04  \\
-2.00/-1.55 &     0.17$\pm$0.04  \\
-2.00/-1.35 &     0.27$\pm$0.04  \\
-2.00/-1.30 &     0.25$\pm$0.04  \\ 
-2.00/-1.25 &     0.27$\pm$0.03  \\
-2.00/-1.10 &     0.30$\pm$0.03  \\
-2.00/-1.05 &     0.29$\pm$0.03   \\
-2.00/-0.95 &     0.30$\pm$0.03  \\
-2.00/-0.90 &     0.31$\pm$0.03
\end{tabular}
\end{table}

On this ground, the least squares solutions performed through our weighted
$<M_V(RR)>$ values with $[Fe/H]<$-1.5 and $[Fe/H]>$-1.5 yield  

\begin{eqnarray}
<M_V(RR)>= & 0.71(\pm0.10)+(0.17\pm0.04)[Fe/H]+ \nonumber \\
& +0.03f \label{eq5}
\end{eqnarray}

\noindent
and 

\begin{eqnarray}
<M_V(RR)>= & 0.92(\pm0.12)+(0.27\pm0.06)[Fe/H]+ \nonumber \\
 & +0.03f, \label{eq6}
\end{eqnarray}

\noindent
respectively, in agreement with the empirical calibrations by Fn93, GS99 
and Cr99 (metal-poor clusters) and MN99 (metal-rich clusters).

\begin{figure}
\vspace{4cm}
\caption{The dependence of $<M_V(RR)>$ on $[Fe/H]$ for the GCs in our 
sample (filled circles: the size of the circle is proportional to the weight 
reported in Table 4) in comparison 
with selected empirical calibrations 
(references are labelled). 
The comparison is performed assuming the minimum (bottom panel) 
and maximum (top panel) 
value of global $Z$ for each given cluster, 
as obtained with the labelled enhancement 
of $\alpha$-elements 
(see text for details). \label{fig2}}
\end{figure}

\begin{figure}
\vspace{4cm}
\caption{As in Fig. 2, but distinguishing between 
Oo I (triangles) and Oo II (circles) globular clusters. The
solid lines are our relations 
for clusters with $[Fe/H]<$-1.5 and 
$[Fe/H]>$-1.5. \label{fig3}}
\end{figure}

\begin{figure}
\vspace{4cm}
\caption{As in Fig. 2, but in the $<M_V(RR)>-\log Z$ plane. 
The different lines show the labelled theoretical calibrations for
$<M_V(ZAHB)>$, as decreased by 0.1 mag to account for the 
evolutionary effects. \label{fig2}}
\end{figure}

As shown in Fig. 3, where OoI and OoII clusters are depicted with 
different symbols, around $[Fe/H]\sim$-1.5 the Oosterhoff 
dichotomy shows off, leading us to guess that the two above  
$<M_V(RR)>-[Fe/H]$ relations hold for the two Oosterhoff groups. More interestingly, 
the same figure suggests that OoII clusters have 
brighter RR Lyraes than OoI clusters with {\it similar} metal 
content, an evidence which coupled with their blue HB 
morphology (see HB types in Table 4) confirms that the 
RR Lyrae evolutionary stage is 
more important than metallicity in triggering the Oosterhoff dichotomy,
{\bf as early suggested by Lee, Demarque \& Zinn (1990) and recently supported
by independent investigations (Lee \& Carney 1999, Clement \& Shelton 1999).}

Figure 4 finally  presents the pulsational $<M_V(RR)>$ results 
as a function of log$Z$, as derived through eq. (4),    
together with selected 
theoretical predictions based on stellar evolution theory. 
The lines drawn in the figure refer to recent $M_V(ZAHB)-\log Z$ 
calibrations as given by Cassisi \& Salaris (1997 [CS]), Cs99, 
Caloi, D'Antona \& Mazzitelli (1997 [CDM]) 
and Ferraro et al. (1999 [Fr99]), with the predicted ZAHB magnitude 
decreased by 0.1 mag to account for the luminosity excess of  
actual RR Lyrae stars over the ZAHB level. 

Inspection of
the figure reveals that none of the theoretical predictions fully 
agrees with our pulsational results with $f$=1, as the 
luminosities provided 
by Cs99 and CDM are systematically too bright, 
while those by CS and Fr99 match our results only 
with log$Z<$-3.0. If $f$=3, then the CS and Fr99 calibrations appear the 
most consistent with our data, but with a tendency to overestimate 
the luminosity of the most metal-rich variables. It seems worth 
noticing that the Cs99 relation is well reproducing all our data 
with $f$=3, but with an overluminosity of about 0.08 mag.

\section{Concluding remarks}

In this paper we have used results from the most recent and
updated computations of non-linear convective pulsating models 
to constrain
the distance modulus of Galactic globular clusters through
the observed periods of $RR_c$ pulsators.
The resulting  $<M_V(RR)>-[Fe/H]$ relation appears in the
range of several empirical linear calibrations, but with
evidence for a non linearity which suggests that the
slope of the relation increases when moving towards 
the metal-richer variables. 
On observational grounds, a similar behavior seems present 
among RR Lyrae stars in $\omega$ Centauri (Rey et al. 2000).  
Moreover, we notice that  over the range of metal-poor stars 
($[Fe/H]<$-1.5) 
our pulsational calibration is in good agreement
with the relations given by Fernley (1993), Groenewegen \& Salaris 
(1999) and Carretta et al. (1999), 
while with $[Fe/H]>$-1.5 it agrees with MacNamara (1999) results. 

Application of our results to RR Lyrae 
stars of the metal-poor globular clusters
in the Large Magellanic Cloud (see data in GS) would give a 
distance modulus of 18.61$\pm$0.12 mag ($f$=1) and 
18.56$\pm$0.12 mag ($f$=3), 
thus supporting the  ``long" distance scale (see also 
Romaniello et al. 1999).

By relying on the present pulsational RR Lyrae absolute 
magnitudes, one derives that the non linearity of 
our $<M_V(RR)>-[Fe/H]$ relation is well reproduced by current predictions 
based on stellar evolution theory. However, in the case of 
solar-scaled chemical compositions, none of the evolutionary predictions
published in the recent literature appears in satisfactory agreement, 
supporting observational evidence for $\alpha$-enhanced chemical mixtures 
in metal-poor stars. With the $\alpha$-elements enhanced by a factor of 3 
with respect to 
iron, the predictions by CS and 
Fr99 agree with our pulsational magnitudes 
even though with a 
tendency of overestimating the luminosity of metal-rich pulsators.  
Interesting enough, one finds that the 
Cs99 relation is well 
reproducing the general dependence of $M_V(RR)$ on log$Z$, 
but with an overluminosity of about 0.08 mag. Holding 
CS99 results, a beautiful agreement with our data would be 
achieved by sistematically increasing the cluster metallicity by 
$\sim$ 0.2 dex, 
an occurrence hardly to be accepted.

To further discuss this point, one has to remind that differences in stellar
models are mainly, if not only, the result of differences in the adopted
input physics.
Discussing RR Lyrae stars in the globular cluster M5 
(Caputo et al. 1999) 
we have already reported pulsational 
evidence suggesting that models with the "most updated"
input physics (as in Cs99) give too luminous HB stars.
Such an evidence has been further supported by independent estimates
based on HIPPARCOS parallaxes for clumping field He burning stars (Castellani
et al. 1999). Data in the previous Fig. 4 reinforce such an evidence, 
suggesting
that the "most updated physics" is probably far from being 
the most adequate one. As a whole, 
we remain with the tantalising evidence that Cs99 models give the rightest 
metal dependence but not right luminosities, whereas those by Fr99 
and CS give
much better luminosities but slightly worst slope.

The role played by the various physical ingredients in
determining the predicted luminosity of HB structures has been recently
discussed in several papers (see Cassisi et al. 1999, Castellani and
Degl'Innocenti 1999, Castellani 1999) and cannot be repeated here. 
However, one may notice that the most recent theoretical predictions
displayed in Fig. 4 all agree within a range of luminosity of about
$\pm$ 0.05 mag. This in our feelings should be regarded as an evidence 
of the high standard reached by evolutionary theories, as well
as a warning that better precision should require a corresponding
level of accuracy in the input physics not yet reached by currently available 
evaluations.

{\bf Acknowledgment:} 

It is a pleasure thank the referee, B. Carney, for his valuable report.
We deeply thank B. Carney and M. Corwin for providing us with  data on NGC 5466 
before publication. Thanks are also due to Santi Cassisi for several warm 
discussions with one of us (F.C.) during an icy week in Teramo.  
Financial support for this work was provided by the 
italian Ministero dell'Universit\`a
e della Ricerca Scientifica e Tecnologica (MURST) under the scientific project
``Stellar Evolution''.

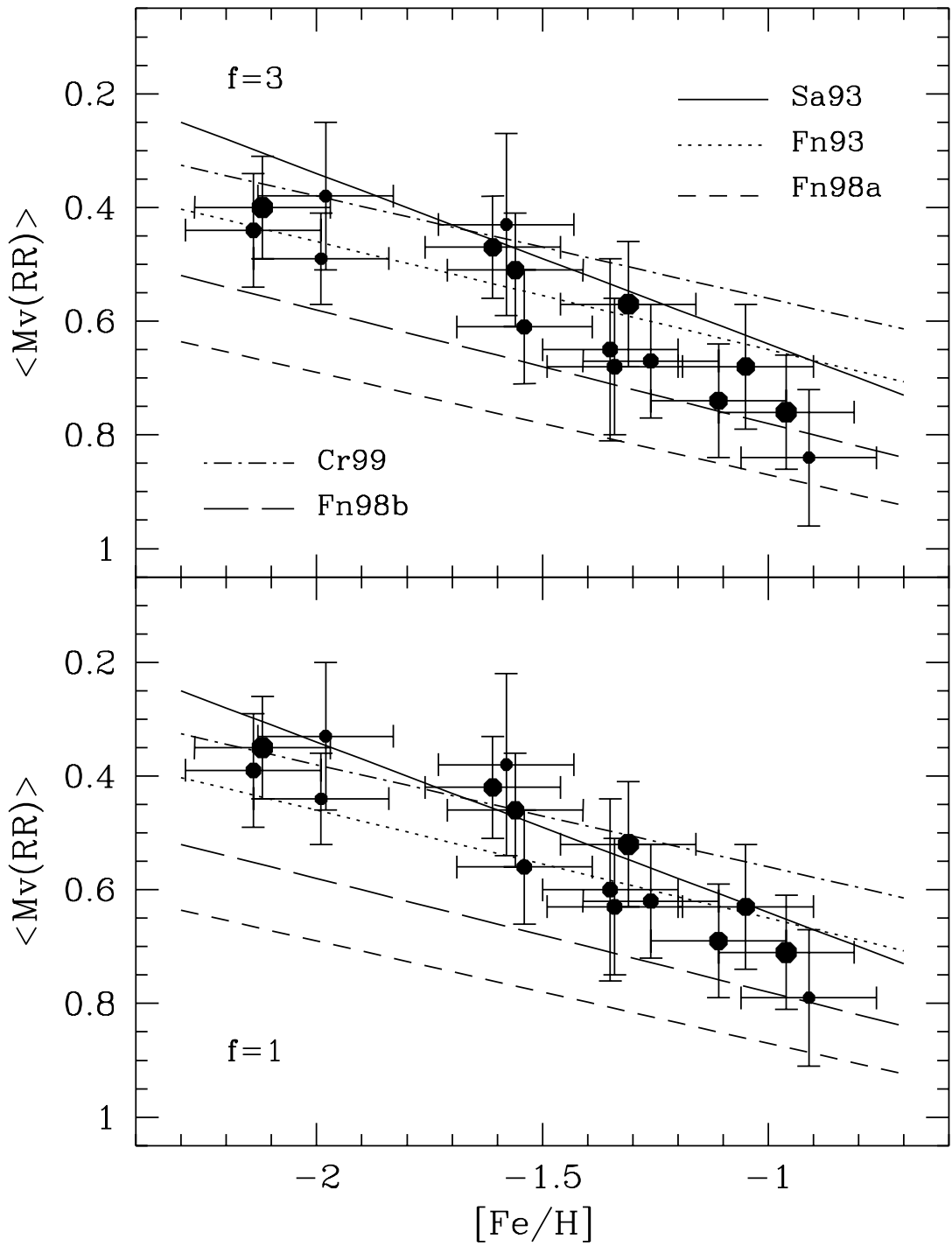
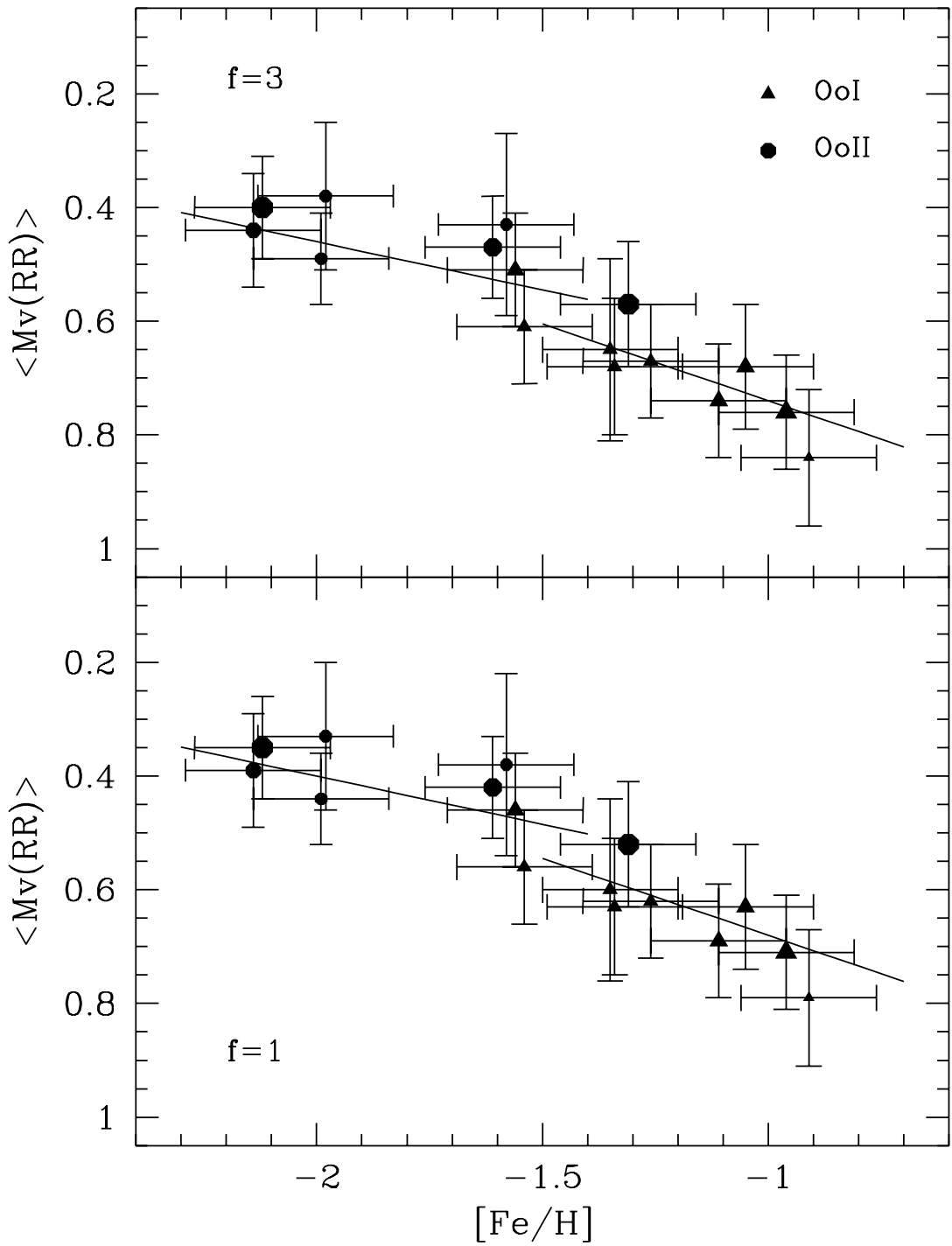
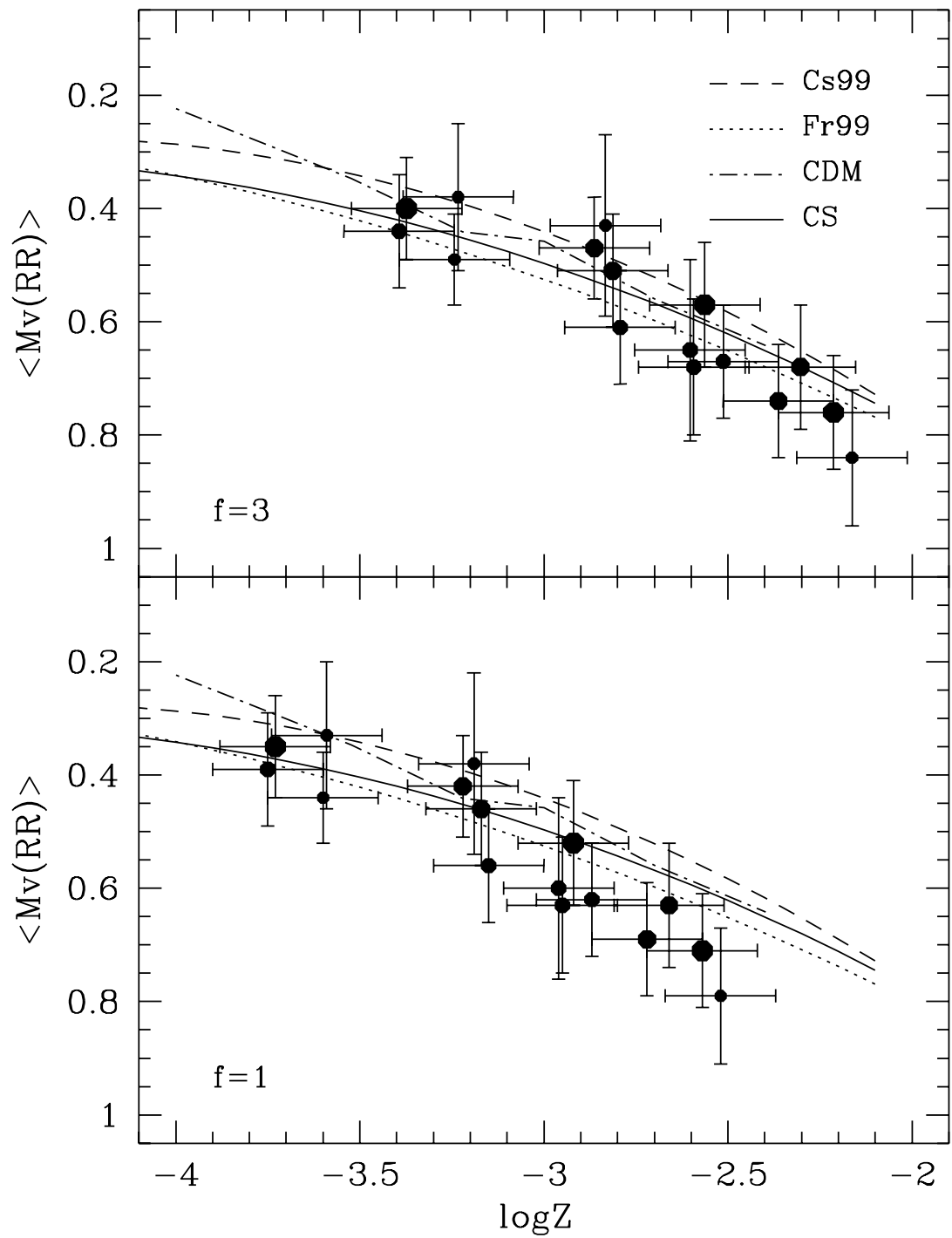
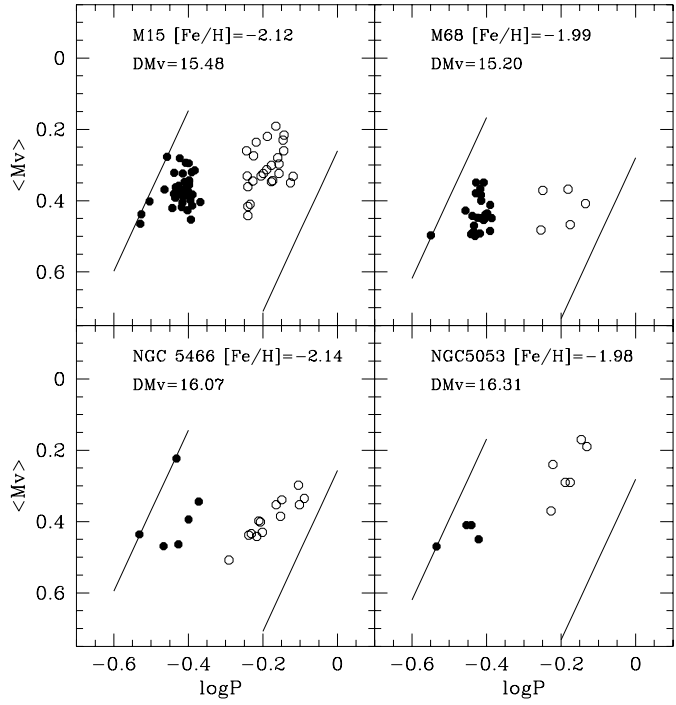
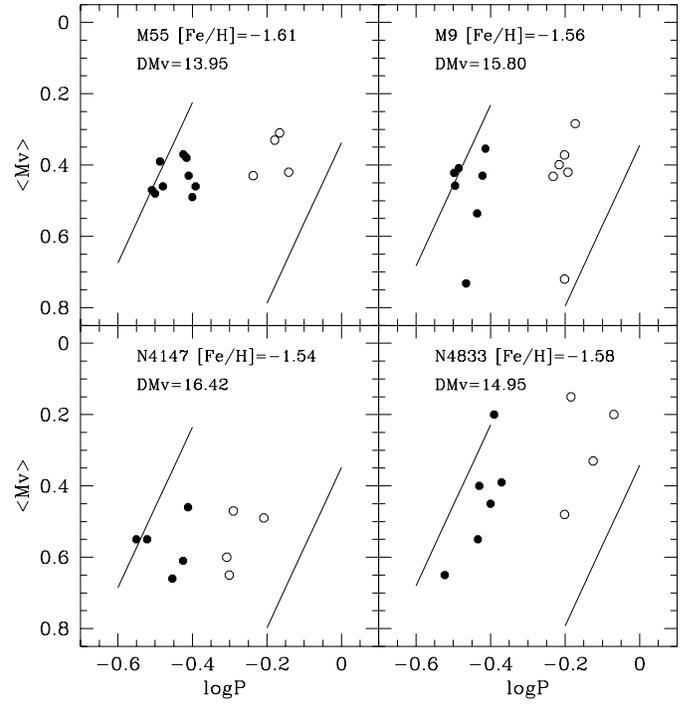
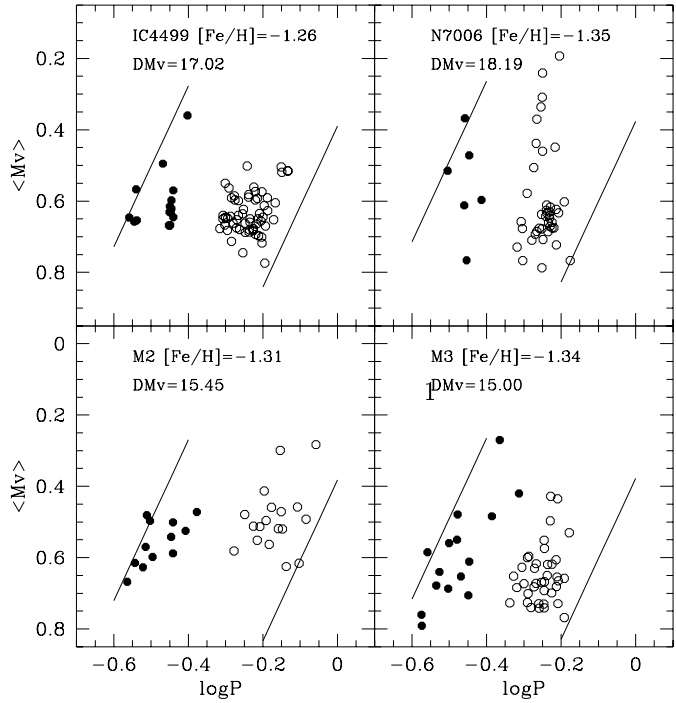
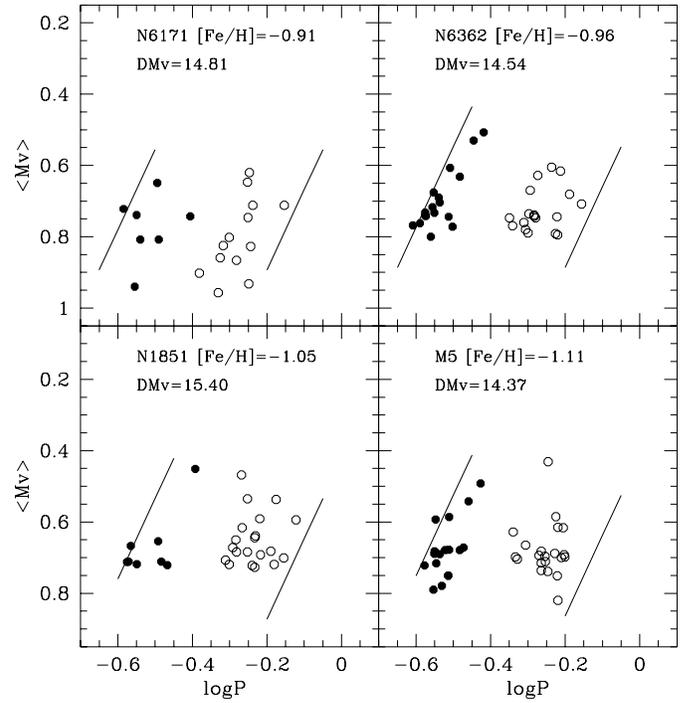
\end{document}